\documentclass[11pt]{article}
\usepackage{amsmath,amsthm}
\usepackage{amssymb}

\usepackage{mathtools}
\usepackage{amsfonts}

\newcommand{\Rtwo}{\overset{2}{R}{}}
\newcommand{\Rthree}{\overset{3}{R}{}}
\newcommand{\gtwo}{\overset{2}{g}{}}
\newcommand{\gfour}{\overset{4}{g}{}}
\newcommand{\gthree}{g}
\newcommand{\COMMENT}[1]{}

\begin{document}


\title{Proof of Positive Energy Theorem \\ by spacetime foliations}

\author{Jacek Jezierski, Piotr Waluk\\
Department of Mathematical Methods in
	Physics, \\ University of Warsaw,
	ul. Pasteura 5, 02-093 Warszawa, Poland\\
E-mail: Jacek.Jezierski@fuw.edu.pl, Piotr.Waluk@fuw.edu.pl}

\maketitle

\section{Introduction}{

Around 1961 R. Arnowitt, S. Deser, and W. Misner proposed, in~their collaborative works, a way of defining ``total four-momentum'' of~a~gravitating system \cite{adm}. The idea consisted in calculating a surface integral, constructed of metric derivatives, at infinity of some spatial hypersurface (``a slice of constant time''). This integral turns out to be well-defined and quite independent of deformations of the chosen hypersurface, as long as the choice is asymptotically flat, i.e., gravitation field falls off quickly enough at infinity.
\begin{displaymath}
\gthree_{ab}=\delta_{ab}+{h_{ab}} \quad h_{ab}\in o(r^{-1})
\end{displaymath}

The energy, or ``mass'', component of ADM four-momentum turned out to be especially useful in various applications. It is given by a~following integral:
\begin{equation*}
M_{ADM}=\lim_{r\to\infty}\frac{1}{16\pi}\oint_{S(r)}(h^j{}_{k,j}-h^j{}_{j,k})dS^k
\end{equation*}

In spite of importance of the concept, it took almost 20 years to settle such basic matter as the question of its positive definiteness.

\vspace{0.3cm}

It was only in 1979 that a complete proof was finally presented by Schoen and Yau, who succeeded by using variational arguments \cite{schoen-yau1},\cite{schoen-yau2}. This method was further generalized to initial data manifolds of dimension up to 7 \cite{schoen-yau3}. Not much later, in 1981, another proof appeared (by Witten), based on the theory of spinors \cite{witten}. An assumption of existence of spin structure was necessary here, but the dimension could be arbitrary.

\vspace{0.3cm}

Here we present a yet alternative approach, requiring only basic tools of differential geometry. This method was conceived by Kijowski and first presented during the IV Marcel Grossman Meeting on General Relativity in Rome (1986). It was later published in a generalized form \cite{positivity}. We summarize these results briefly and present a new foliation gauge.

\phantom{o}
\vspace*{-0.2cm}
}

\section{Tools}{

Two noteworthy theorems play an important role in our reasoning:

\subsection{Gauss-Codazzi equations}{

The Gauss-Codazzi equations express a relation between the geometry of a hyper-surface and its enveloping space.
\begin{equation}
\begin{aligned}
\overset{d}R{}^m{}_{ijk} &= \overset{d-1}R{}^m{}_{ijk} \pm (K_{ij}K_k{}^m-K_{ik}K_j{}^m) \\
\overset{d}R{}^\mu{}_{ijk}n_\mu &= \mp (K_{ij|k}-K_{ik|j}) \label{gauss-codazzi}
\end{aligned}
\end{equation}
Symbols $\pm$ and $\mp$ above correspond to the sign of $n_\mu n^{\mu}$.

\vspace*{0.5cm}

These mathematical identities give rise to constraints on initial data $(\gthree_{ab},K_{ab})$ in the Cauchy problem for Einstein equation:
\begin{align}
16\pi T_{\mu\nu}n^\mu n^\nu&=16\pi\rho=\Rthree+K^2-K_{ab}K^{ab} \label{scalarConstraint}\\
8\pi T_{a\mu}n^\mu&=K^b{}_{a|b}-K_{|a} \notag
\end{align}

}

\subsection{Gauss-Bonnet theorem}{

For any 2-dimensional surface $P$, the following equality holds:
\begin{equation}
\int_{P}\frac 12 R(P)dA+\int_{\partial P}k_g ds=2\pi\chi(P)
\label{GaussBonnet}
\end{equation}
where $R(P)$ is the scalar curvature of $P$, $k_g$ denotes geodesic curvature of the boundary, and $\chi(P)$ is the Euler characteristic of the surface, a topological invariant.

}

}

\section{Preliminary assumptions}

We are considering a four-dimensional space-time manifold, equipped with a metric of Lorentzian signature $(M, \gfour_{\mu\nu})$. We assume that it is globally hyperbolic and, furthermore, \emph{allows a maximal Cauchy surface} (i.e. a spatial hypersurface for which the trace of extrinsic curvature vanishes).

All matter fields are presumed to satisfy the weak energy condition:
$T_{\mu\nu}X^\mu X^\nu	= \rho \geq 0$ for $X^\mu$ - timelike vector field.

Our maximal Cauchy surface $\Sigma$ is supposed to be asymptotically flat (a necessary condition for defining the ADM mass). We study the simplest case - $\mathbb{R}^3$ topology - only, but the result should be easy to generalize.

Finally, we demand existence of a foliation of $\Sigma$ with two-dimensional leaves, satisfying a certain ``gauge condition''.

\subsection{Notation remarks}

We choose the coordinate system in a way coinciding with our choice of $\Sigma$ and its intrinsic foliation. That is, the Cauchy surface is the level set of the time coordinate and (2+1) foliation leaves are defined by one of the spatial coordinates (as level sets of $\tau(x^i):=x^3$). The following index convention is used:

\begin{itemize}
	\item $\mu, \nu$ - small Greek indices run over all coordinates on $M$ $(0,1,2,3)$
	\item $a, b$ - small Latin indices enumerate coordinates on $\Sigma$ $(1,2,3)$
	\item $A, B$ - big Latin indices denote coordinates on 2-dimensional leaves $(1,2)$
\end{itemize}

\vspace*{-0.4cm}
Metric tensors for the 4, 3, and 2-dimensional geometries are $\gfour_{\mu\nu}$, $\gthree_{ab}$ and $\gtwo_{AB}$, the~latter two being induced by the former.
Covariant derivatives associated with $\gfour_{\mu\nu}$ and $\gthree_{ab}$ are denoted \mbox{by $;$ and $|$ (or $\nabla$),} respectively.
$K=-n_{a;b}$ is the extrinsic curvature (second fundamental form) of the Cauchy surface and $K=K_{ab}\gthree^{ab}$ its trace ($n^\mu$ being the unit normal field to $\Sigma$).
Analogously: $M^{a}$, $k_{AB}=-M_{A|B}$, $k=k_{AB}\gtwo^{AB}$ are the unit normal field, extrinsic curvature, and its trace for 2-dimensional foliations of $\Sigma$.
In addition: $a^i:=M^jM^i{}_{|j}$, $\phi:=\ln \gthree^{33}$, and $\lambda=(\det \gtwo_{AB})^{1/2}$ - volume form on foliation leaves.

\section{The energy formula}

The general idea of the proof is straightforward. We wish to express the ADM mass as a sum of non-negative integrals. The main step to achieve this consists of combining the scalar constraint equation \eqref{scalarConstraint} with Gauss-Codazzi identities for 2-dimensional foliations \eqref{gauss-codazzi} to obtain, after some rearrangements:
\begin{equation}
\label{hamiltonian}
\begin{aligned}
2\partial_i (\lambda M^i k + \lambda a^i )+\lambda \Rtwo =
16 \pi \rho \lambda & +\lambda K_{ab}K^{ab}  +\lambda(k_{AB}k^{AB}-\frac 12 k^2) + \\
+ \lambda \gtwo{}^{AB}\partial_A\phi\partial_B\phi
&\underbrace{+ \lambda k M^i \partial_i \phi - \frac 12 \lambda k^2}_{\hspace*{-3cm}\mathclap{ \substack{ \text{Controlled by an appropriate} \\ \text{choice of gauge condition}}}}
\underbrace{-\lambda K^2 \vphantom{\frac 12}}_{\mathclap{\qquad\qquad\qquad\qquad =\: 0 \text{ by maximality of } \Sigma}}
\end{aligned}
\end{equation}

The resulting equation plays a key role in our proof. Its right hand side consists mostly of manifestly positive terms, apart from the last three. One of these is actually equal to zero due to maximality of $\Sigma$, but we include it in the above formula to keep it general.
The other two must be dealt with by imposing a ``gauge condition'' on the (2+1) foliation. An appropriate choice will allow us to rewrite the right hand side with positive terms only, possibly introducing some small changes to the left hand side as well in the process. Various conditions are available for consideration and two of them are discussed further on.

Once the right hand side is re-expressed in a manifestly positive way all that is left is to show that the left hand side will, when integrated over $\Sigma$, yield the ADM mass multiplied by some positive factor. Some further remarks on that matter are given in one of the following sections.

\section{$\beta$ - foliations}{

The ``$\beta$-foliations'' are foliations that satisfy gauge equations proposed and discussed in works of J.Jezierski and J.Kijowski \cite{positivity},\cite{betafoliations}. Two topological situations have been considered - flat and spherical. We summarize them briefly:

\subsection{Topologically flat}{

Assuming $\Sigma$ to have the topology of $\mathbb{R}^3=\mathbb{R}^2\times\mathbb{R}^1$ we demand that the following gauge condition be~satisfied:
\begin{equation}
\label{betaflatgauge}
w_\beta=k-\beta M^i\partial_i \phi = 0
\end{equation}
Applying this to the troublesome terms in \eqref{hamiltonian}:
\begin{equation*}
\lambda k M^i \partial_i \phi - \frac 12 \lambda k^2 = \frac 12 \lambda\beta(2-\beta)(M^i\partial_i \phi)^2
\end{equation*}
we get a manifestly positive expression for $\beta\in [0,2]$.
Rewriting the gauge condition as a differential equation for $\tau$, we get:
\begin{equation}
\label{betaflatequation}
(\det \gthree_{ab})^{1/2}\nabla_i(|\nabla \tau|^{2\beta-1}\nabla^i \tau)=0
\;\; \iff \;\;
\delta\int_{\Sigma} (\det \gthree_{ab})^{1/2}|\nabla \tau|^{2\beta+1}\mathrm{d}^3x=0
\end{equation}
Some special cases are noteworthy here:
\begin{itemize}
	\item For $\beta=0$ the condition \eqref{betaflatgauge} becomes $k=0$, i.e. minimality condition for foliation surfaces,
	\item For $\beta=\frac 12$ equation \eqref{betaflatequation} becomes a laplacian $\Delta \tau = 0$,
	\item For $\beta=1$ equation \eqref{betaflatequation} becomes conformally invariant.
\end{itemize}

}

\subsection{Spherical}{
We consider $\Sigma \backslash \{ x_0\}=\mathbb{S}^2\times\mathbb{R}_{+}$, with $x^3$ being the radial coordinate and foliation leaves having the~topology of 2-spheres. The gauge condition is now stated as:
\begin{equation*}
w_\beta = k-\beta M^i\partial_i\phi+2|\nabla r|/r = 0
\end{equation*}
This allows us to rewrite:
\begin{equation*}
\lambda k M^i \partial_i \phi - \frac 12 \lambda k^2  = - 2\lambda(1-\beta)M^3 M^i(\partial_i\phi)/r - \frac{2\lambda \gthree^{33}}{r^2} + \frac 12 \lambda\beta(2-\beta)(M^i\partial_i \phi)^2
\end{equation*}

Transferring the first two terms to the left side of \eqref{hamiltonian}, we
once again obtain manifest positivity of its right hand side for $\beta\in [0,2]$.

The corresponding differential equation has the form:
\begin{equation*}
(\det \gthree_{ab})^{1/2}\nabla_i(r^{-2}|\nabla r|^{2\beta-1}\nabla^i r)=0
\quad \xRightarrow[\substack{\phantom{Q}\\ \hspace*{-1.0 cm}\rho:=\frac{\beta}{\beta-1}(r^{(\beta-1) / \beta}-1) \hspace*{-1.0 cm}}]{} \quad
\nabla_i(|\nabla \rho|^{2\beta-1}\nabla^i \rho)=0
\end{equation*}
The equation for $\rho$ looks the same as \eqref{betaflatequation}, but it has different  boundary conditions.

It is noteworthy that $\beta=0$ corresponds to a condition for extrinsic curvature \linebreak[4]
\mbox{$(\det \gthree_{ab})^{1/2}k=-\lambda/r$,} satisfied, for example, by radial coordinates of Schwarzschild and Reissner-Nordstroem metrics, written down in Schwarzschild coordinates.

}

}

\section{$\alpha$ - foliations}{

We state the following gauge condition:
\begin{equation}
\label{alphagauge}
M^i \partial_i k + \alpha k^2 = \frac 12 k M^j \partial_j \phi
\quad \xLeftrightarrow[\text{for } k \neq 0]{} \quad
\frac{\gthree^{3i}}{\gthree^{33}} \partial_i \Big( \frac{(\gthree^{33})^{1/2}}{k}\Big)=\alpha
\end{equation}
which allows us to rewrite \eqref{hamiltonian} in the form:
\begin{equation*}
(\lambda M^i k + 2\lambda a^i )_{,i} + \lambda \Rtwo = \text{positive terms}
+ (\frac 12 + \alpha)\lambda k^2
\end{equation*}
yielding manifest positivity of the right hand side for $\alpha\geq-\frac 12$.\\
This condition can also be expressed as a differential equation for $\tau$:
\begin{equation*}
\frac{\nabla^i \tau}{|\nabla \tau|^2}\partial_i \Big(\frac{|\nabla \tau|}{\nabla^j (\frac{\nabla_j \tau}{|\nabla \tau|})} \Big) = \alpha
\end{equation*}
\begin{equation*}
(\nabla^i \tau)\partial_i \Big(\frac{\nabla^j (\frac{\nabla_j \tau}{|\nabla \tau|})}{|\nabla \tau|} \Big) + \alpha \big(\nabla^j (\frac{\nabla_j \tau}{|\nabla \tau|})\big)^2 = 0
\end{equation*}
To get a feeling of its behaviour in the asymptotically flat region, we analyse solutions in~flat Euclidean space. Those are quite abundant, for example:

\begin{itemize}
	\item Foliation of $\mathbb{R}^3$ with flat planes obviously satisfies \eqref{alphagauge} for any $\alpha$, as does any foliation with minimal surfaces of $\gthree_{ab}$.
	\item In a spherical coordinate system the radial coordinate yields $\alpha=-\frac 12$, while $\tau = \sin ^p \theta$ is a solution corresponding to $\alpha=-p$
	\item $\rho^p$, $\rho$ being the radial coordinate in the cylindrical system, is a solution with $\alpha=-p$
	\item $\tau = \cosh \mu$, gives $\alpha=\frac 12$. $\mu$ here is one of the bispherical coordinates $(\sigma,\mu,\phi)$:
\begin{equation*}
x=\frac{a \sin \sigma \cos \phi}{\cosh \mu - \cos \sigma} \quad
y=\frac{a \sin \sigma \sin \phi}{\cosh \mu - \cos \sigma} \quad
z=\frac{a \sinh \mu}{\cosh \mu - \cos \sigma}
\end{equation*}
	\item $\tau = \cos \sigma$, gives $\alpha = -\frac 12$, when $\sigma$ is one of the toroidal coordinates:
\begin{equation*}
x=\frac{a \sinh \mu \cos \phi}{\cosh \mu - \cos \sigma} \quad
y=\frac{a \sinh \mu \sin \phi}{\cosh \mu - \cos \sigma} \quad
z=\frac{a \sin \sigma}{\cosh \mu - \cos \sigma}
\end{equation*}

\end{itemize}

It is also noteworthy that the radial coordinate of both \mbox{Reissner-Nordstroem} and Schwarzschild is a solution with $\alpha = -\frac 12$.

}

\section{Integration of the ADM mass}

By assuming that the metric on $\Sigma$ behaves asymptotically in the following way:
\begin{equation*}
\gthree_{ab}=\Big(1+\frac {M_{ADM}}r + o(r^{-1})\Big)\gamma_{ab}, \quad \gamma_{ab}=\delta_{ab} + o(r^{-1}), \quad \partial_c \gamma_{ab} = o(r^{-2}),
\end{equation*}
one can integrate the left hand side of \eqref{hamiltonian} (possibly with modifications introduced by the foliation gauge) over the Cauchy surface and show that it evaluates to $16\pi M_{ADM}$ for both cases of $\beta$-foliations.
For spherical $\beta$-foliations a slightly modified example may be considered, with a ball cut out at the origin instead of o single point. The volume integral then can be rewritten as a difference of two surface integrals - one at infinity, again giving $16\pi M_{ADM}$, and a second one, over the surface of the ball. This second integral might have an interpretation of black hole mass in some situations and vanishes as the radius of the ball tends to zero, as long as the metric is non-singular.

Evaluation of this integral in both cases requires the use of Gauss-Bonnet theorem \eqref{GaussBonnet}, which makes the derivation dependent on the topology of foliation leaves. This poses a problem when we try to analyse $\alpha$-foliations in the general case, where the topological setting is not known. If one assumes, however, an analogous situation to one of those considered for $\beta$-foliations, then the integral of our interest can be evaluated in a similar way. Modifications in the left hand side, introduced by the $\alpha$-foliation gauge, effect there in a different factor multiplying the ADM mass.

We refer the reader to \cite{bachelor} and \cite{positivity} for the detailed procedure and calculations.

\section{Final remarks}{
\begin{itemize}
	\item Vanishing of the integral of the left hand side of (appropriately gauged) equation \eqref{hamiltonian} implies vanishing of all the right hand side terms on $\Sigma$. By uniqueness of initial data evolution we thus see that vanishing of the ADM mass implies emptiness ($T_{\mu\nu}=0$) and flatness of the whole space-time.
	\item Gauge conditions presented above also find application in examining the degrees of freedom of gravitational field \cite{betafoliations}.
	\item As most of the reasoning presented above does not actually depend on the dimension of the Cauchy surface it might be possible to generalize the proof to higher-dimensional spacetimes.
\end{itemize}
}


\end{document}